\documentclass{iau}

\usepackage{amsmath}
\usepackage{graphicx}
\usepackage{multirow}

\begin{document}

\lefttitle{Ansar et al.}
\righttitle{IAU Symposium 379: Template}

\jnlPage{1}{7}
\jnlDoiYr{2023}
\doival{10.1017/xxxxx}

\aopheadtitle{Proceedings of IAU Symposium 379}
\editors{P. Bonifacio,  M.-R. Cioni \& F. Hammer, eds.}

\title{Dark Matter Halo Spin of the Dwarf Galaxy UGC 5288: Insights from Observations, N-body and Cosmological Simulations}

\author{Sioree Ansar$^{1,2^*}$, Sandeep Kumar Kataria$^{3,4}$\& Mousumi Das$^1$}
\affiliation{$^1$ Indian Institute of Astrophysics, Bangalore 560034, India, \\
$^2$ Pondicherry University, R.V. Nagar, Kalapet 605014, Puducherry, India \\
$^3$ Department of Astronomy, School of Physics and Astronomy, Shanghai Jiao Tong University, 800 Dongchuan Road, Shanghai 200240, China\\
$^4$ Key Laboratory for Particle Astrophysics and Cosmology (MOE) / Shanghai Key Laboratory for Particle Physics and Cosmology, Shanghai 200240, China \\
$^{*}$\email{sioreeansar@gmail.com}}

\begin{abstract}
Dark matter (DM) halo angular momentum is very challenging to determine from observations of galaxies. In this study, we present a new hybrid method of estimating the dimensionless halo angular momentum, halo spin of a gas-rich dwarf barred galaxy UGC5288 using N-Body/SPH simulations. We forward model the galaxy disk properties- stellar and gas mass, surface densities, disk scalelengths, bar length and bar ellipticity from observations. We use the HI rotation curve to constrain the DM halo density profile and further use the bar properties to determine the models that best represent the observed baryonic disk. We compare the halo spin profile from our models to the halo spin profiles of similar mass dwarf galaxy analogues of UGC5288 in the TNG50 simulations. The halo spin profile from our simulated models matches within ballpark values of the median spin profile of UGC5288 analogues in the TNG50 simulations, although there are some uncertainties due to the DM halo evolutionary history.
\end{abstract}

\maketitle
\section{Introduction}
One of the current challenges in the determination of DM halo properties is to estimate the halo angular momentum. The DM halo angular momentum is difficult to measure for observed galaxies because of the lack of probes to measure the motion of the DM. In simulations the DM halo spin $\lambda$ represents a dimensionless measure of the halo angular momentum \citep{Peebles.1969}.
\begin{equation} \label{eq:spin}
        \lambda  = \frac{J |E|^{1/2}}{G M^{5/2}}
\end{equation}
where $J$ is the total angular momentum, $E$ is the total energy and $M$ is the mass within a radius $r$. Here, we investigate a new hybrid method combining semi-analytical techniques and N-Body/SPH simulations to forward model a dwarf Low Surface Brightness barred galaxy UGC 5288, to determine its DM halo density profile and spin profile. We consider two major selection criteria: (1) an isolated galaxy with no signatures of interaction with other galaxies in the recent past, as we want to model an isolated system and (2) the galaxy should have very low star formation, as we do not model star formation in our simulations. 

\section{Galaxy modelling}
\begin{figure}[h]
\centering
    \includegraphics[scale=.19]{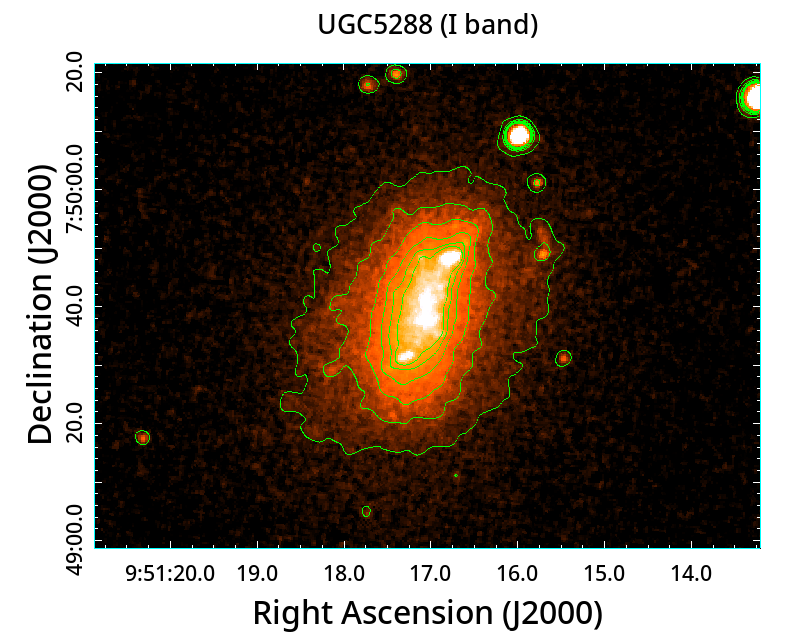}
  \includegraphics[scale=.3]{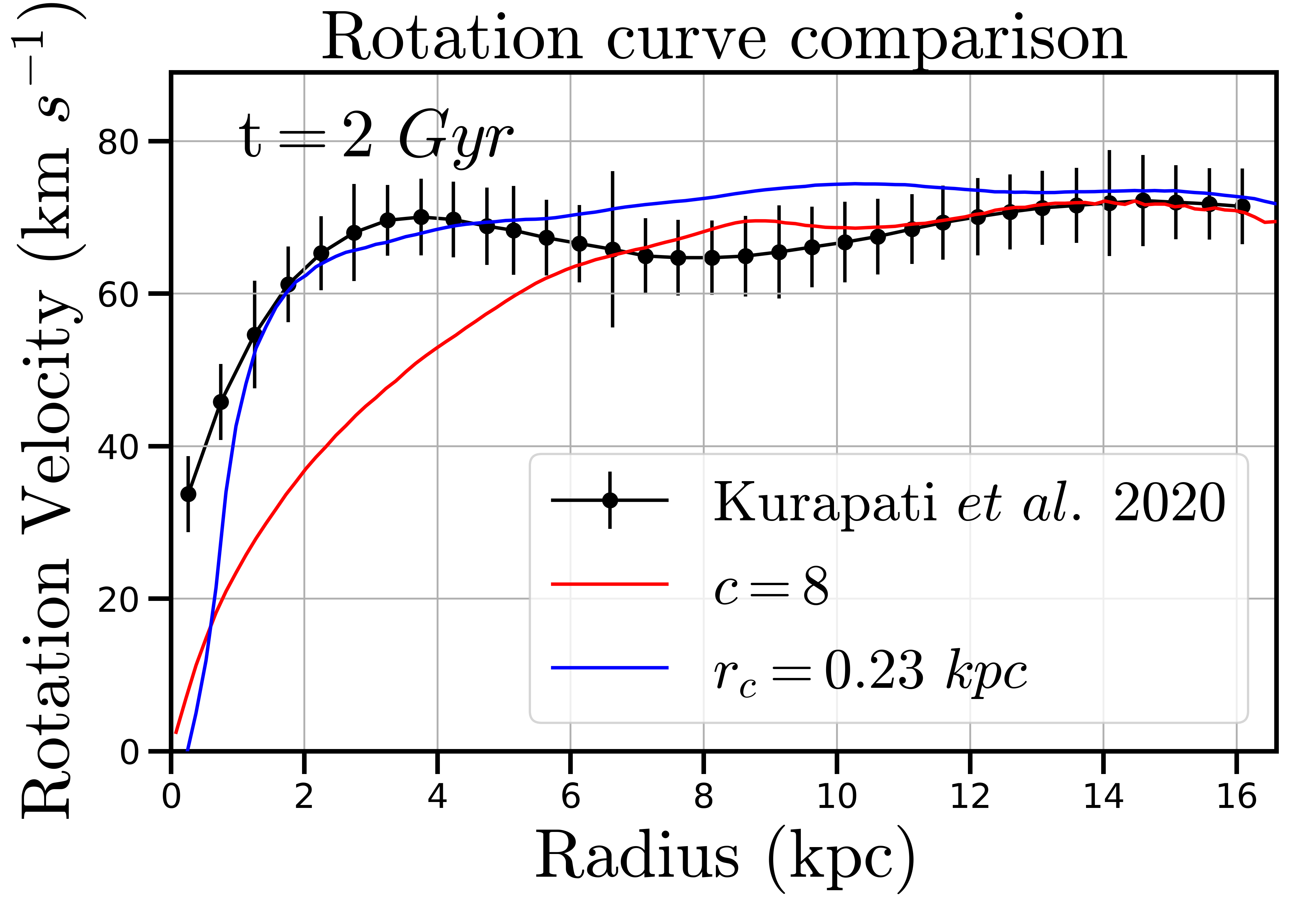}
  \caption{I-band image of UGC5288 (panel (a)). The contours show 0.6, 0.5, 0.4, 0.3, 0.2 and 0.1 times the peak flux in the image and highlight the barred structure. Panel (b) shows the comparison of the gas rotation curves of the best fit simulated models of Hernquist with halo concentration $c=8$ (red line) and PISO halo profiles with core radius $r_{c}=0.23$ kpc (blue line), with the HI rotation of UGC5288 (black line) from \cite{Kurapati.et.al.2020}. The PISO models show a better match to the observed HI rotation curve. }
  \label{fig:rotation_curves}
\end{figure}

UGC5288 is a classical dwarf galaxy in the Lynx cancer void. It has an extended, undisturbed gas disk without signs of any interaction with other galaxies, a very low star formation rate (\cite{werk.et.al.2011}) and a well-studied HI rotation curve \citep{Kurapati.et.al.2020}. We forward model UGC 5288's disk properties (stellar mass, gas mass, stellar and gas disk scalelengths and bar properties) and HI rotation curve in N-body/SPH simulations (see Figure \ref{fig:rotation_curves}). We model UGC 5288 with a stellar disk and gas disk with exponential profiles in radial direction and $sech^{2}$ profile in the direction perpendicular to the disk plane. We model the DM halo of the galaxy using two models- the cuspy Hernquist halo profile and the flat-core pseudo-isothermal (PISO) halo model. We use the N-body simulation code \textsf{GalIC} \citep{Yurin.and.Springel2014} to generate the initial conditions and \textsf{GADGET2} \citep{Springel.2005} to evolve the galaxy for a few Gyrs to ensure the stability of our simulated models. For more details about the modelling of the galaxy and the simulations see \cite{Ansar.et.al.2023}. \\
\vspace{-0.2cm} \\
 \hspace*{0.4cm}  (a)  \hspace{5.9cm}    (b) \\
 \vspace{-1cm}
\section{Essential features of Dark matter halo modeling}
Here we discuss in detail the salient features in the galaxy DM halo modeling.
\begin{figure}[h]
\centering
  \includegraphics[width=0.9\textwidth]{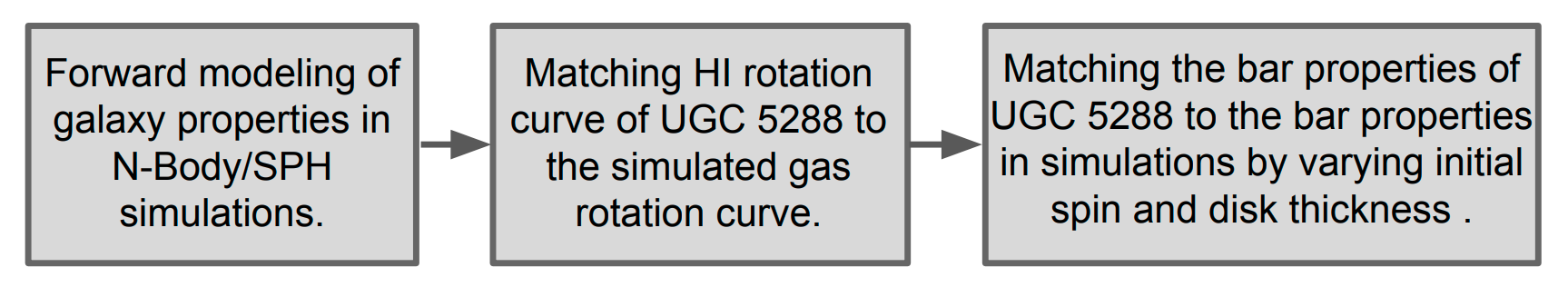}
  \caption{Flowchart of the method to estimate the halo spin profile of UGC5288. }
  \label{fig:flowchart}
\end{figure} \\
\noindent
{\it Importance of the steps of modelling.} After matching the observed properties of UGC5288 with the stable simulated models, constraining the DM halo profile with the HI rotation curve is an essential step in our method to finally estimate the halo spin profile. Following the flowchart in Figure \ref{fig:flowchart} (also see \cite{Ansar.et.al.2023} for more details), first we need to constrain the dark matter halo profile (using the HI rotation curve) and in the next step, to compare the observed bar properties (bar position angle and ellipticity profiles) with the bar properties in the simulations. The reason is that the predicted spin profile is sensitive to the initial DM density profile (also pointed out in Fig. 11 of the \cite{Ansar.et.al.2023}). This dependence is expected as the halo spin (Equation \ref{eq:spin}) has a strong dependence on the total DM mass inside a given radius. Hence we need to constrain the gas rotation curve in the simulations so that the underlying DM density profile is constrained. After the initial estimation of the DM profile from the rotation curve, the possibility of varied halo spin distribution is reduced a bit. We note that other assumptions of the velocity structure of the DM particles will also affect the final halo spin distribution. So we consider the most general velocity structure of the DM halo for our models with three integrals of motion in modelling. Very few of these models (based on the initial halo spin parameter and disk scale height) ultimately give a disk that hosts a weak bar. 
Now, what happens if we first constrain the bar properties? If we do not constrain the DM profile and in the first step try to constrain the bar properties, then there will be much more scatter in the values of possible halo spin. So after constraining the DM mass distribution, we compare the bar properties of our simulated models with the observed bar in UGC 5288. This reduces the scatter in the halo spin distribution. This is the best we could do at this moment as we lack information about the evolutionary history of the galaxy we are trying to model. 

\noindent
{\it Presence of a stellar bar.} We also test the impact of the presence of a bar in stellar disks hosted by DM halos with a similar type of mass distribution. We select barred and unbarred galaxies in the TNG50 and TNG100 data sets in different stellar mass ranges, having similar dark matter circular velocity curves. We show that the median spin values are different for the barred and unbarred galaxy groups even though the DM circular velocity curves are very similar (see Figure L.1. in \cite{Ansar.et.al.2023}). This shows that the presence of a bar affects the halo spin irrespective of the effect of the dark matter density distribution. We are studying this in more detail in a following article (Ansar et al. in prep).

\noindent
{\it Can the method be used for unbarred galaxies?} The barred structure will introduce an asymmetric potential and it will interact with the spherical dark matter halo by exchanging angular momentum. Hence in principle, it should significantly alter the final spin distribution and deviate it from that of an unbarred galaxy. In support of this argument, the connection between halo spin and bar properties has been already discussed in various idealized N-body simulations \citep{Long.et.al.2014,Collier.et.al.2018,Kataria.Shen.2022}, as well as TNG50 simulations \citep{Rosas-Guevara.et.al.2022}, where the authors show that bars are more prominent in low-spin halos compared to high-spin dark matter halos. With a barred galaxy, we have a small range of values of halo spin that helps in recreating the baryonic disk properties. So it helps us narrow down on some specific models. If the galaxy is unbarred, we will only be using the constraints from the rotation curve for the DM mass distribution, but then we will have a different and wider range of values for halo spin and initial disk thickness that creates an unbarred galaxy. Hence, we will get more scatter in the values of halo spin for modelling an  unbarred galaxy.

\section{Comparison with cosmological simulations}
\begin{figure}[h]
\centering
  \includegraphics[scale=.23]{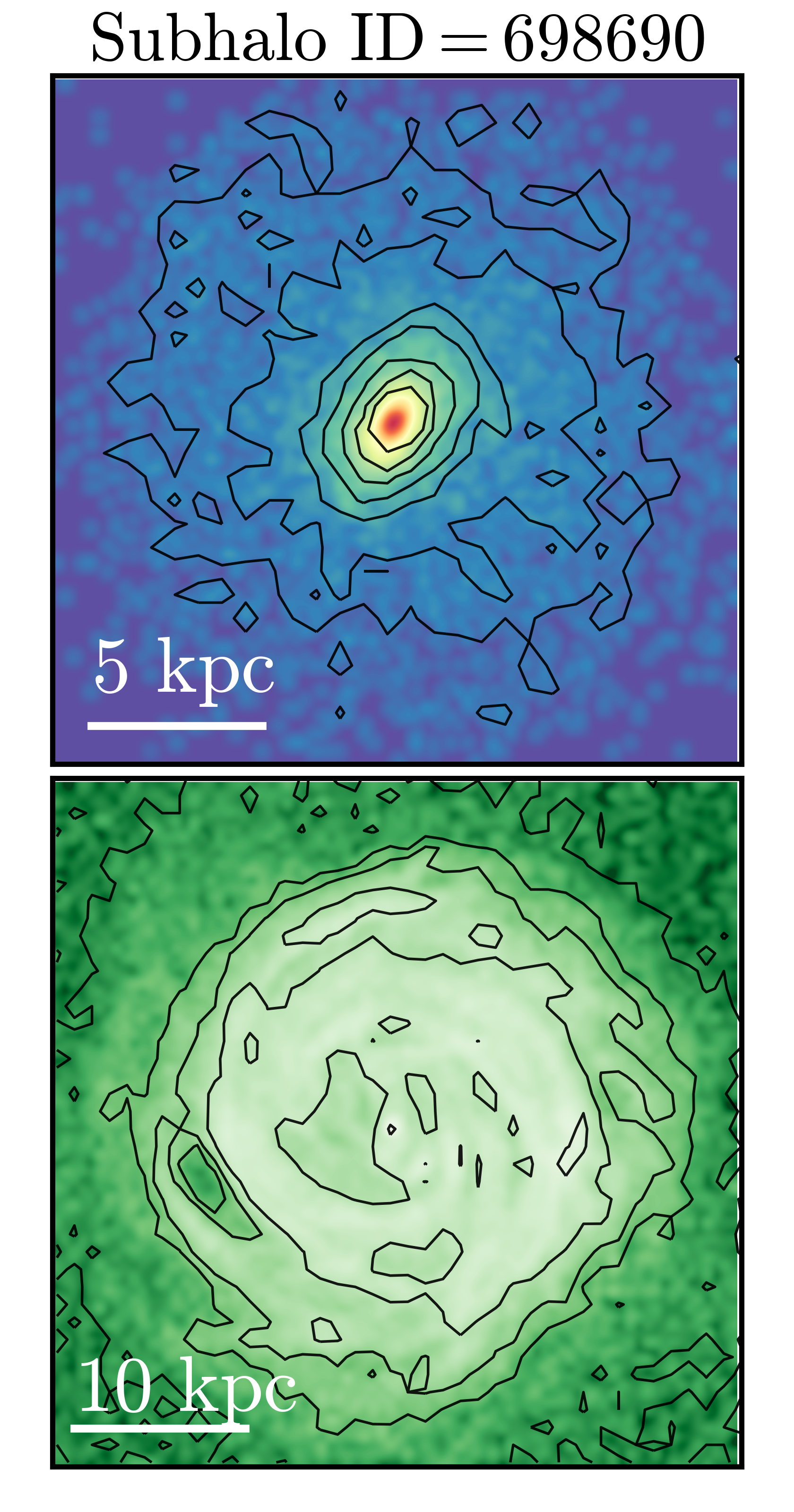}
  \includegraphics[scale=.23]{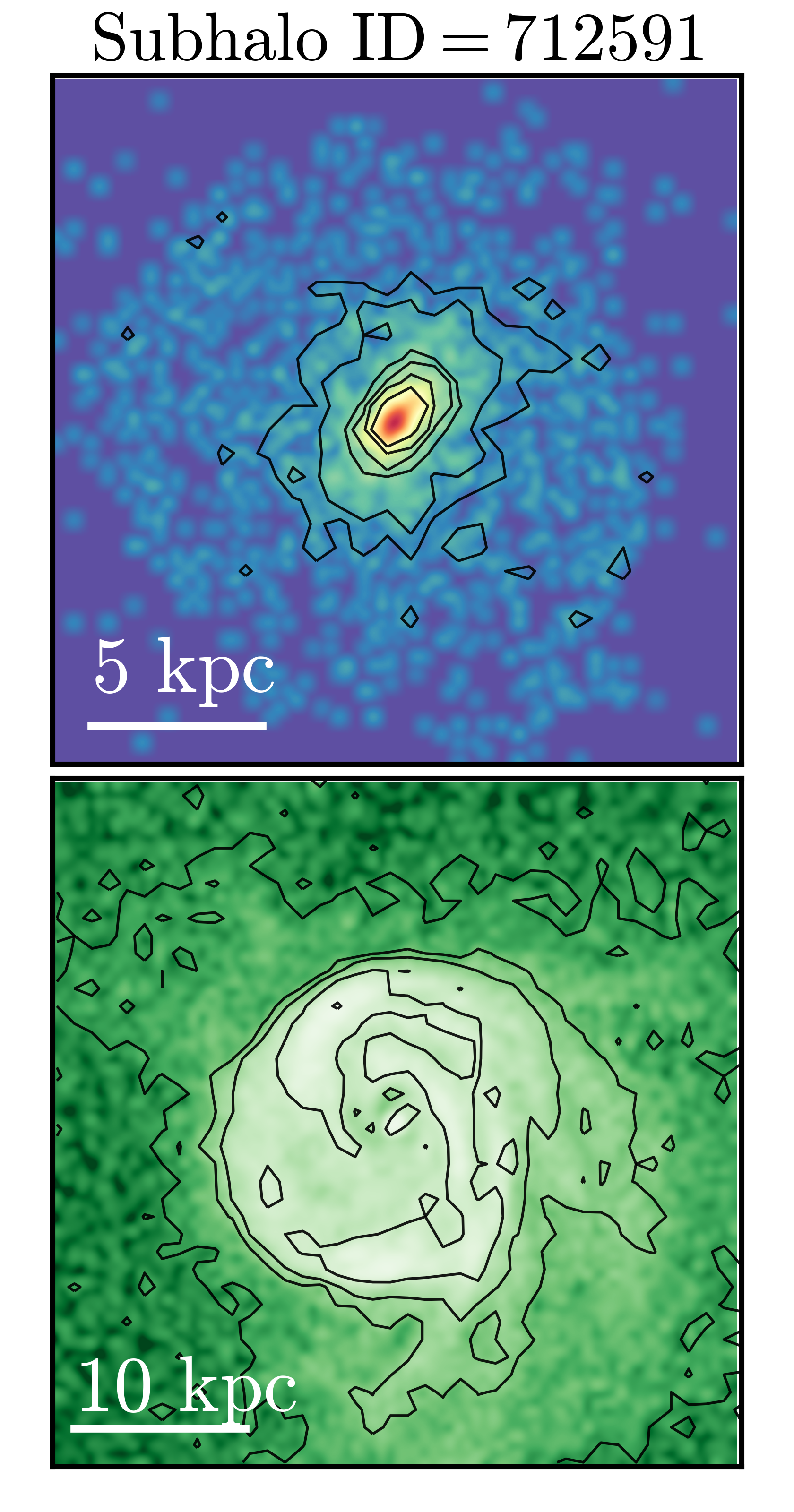}
\includegraphics[scale=.23]{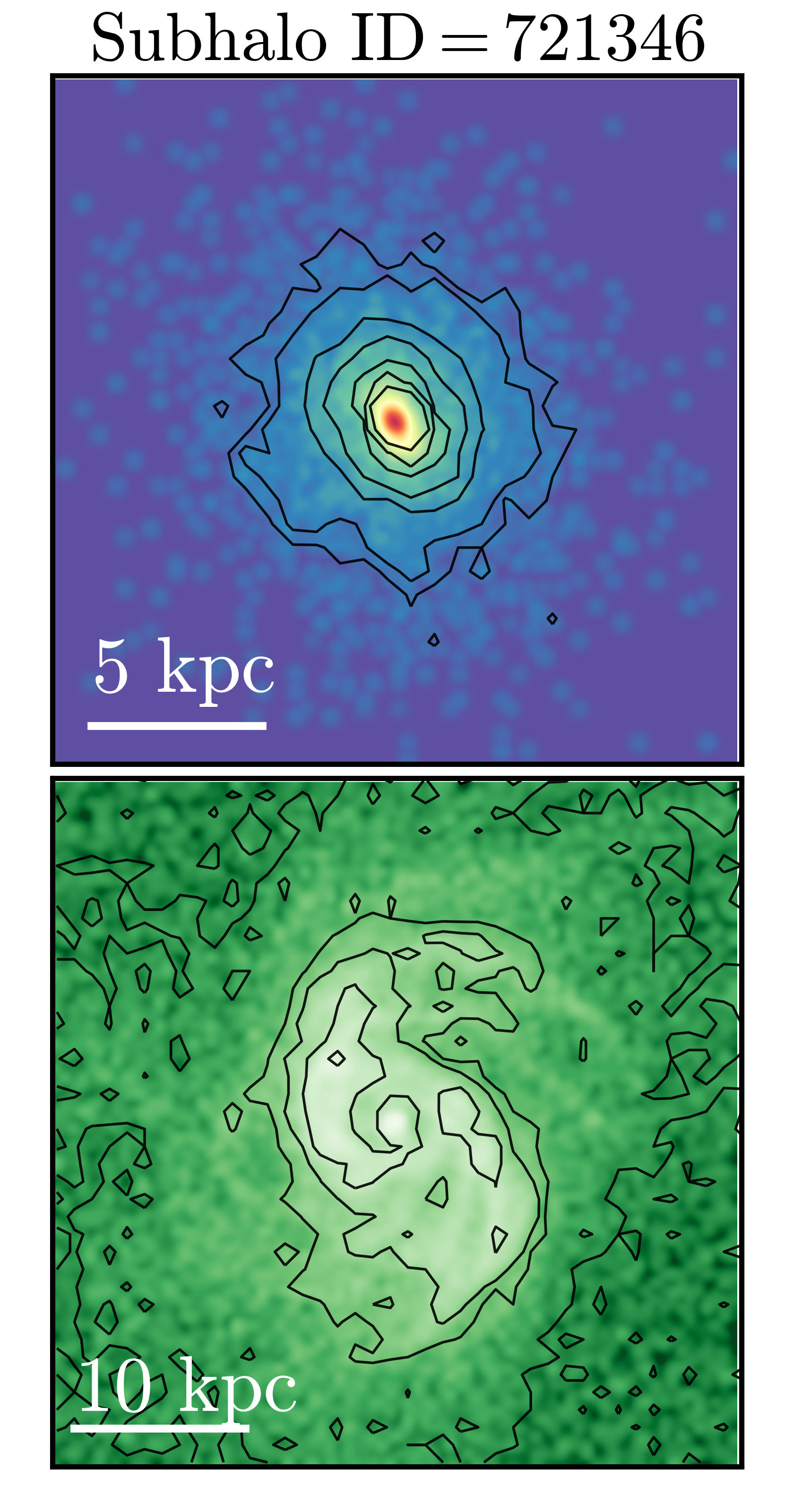}
  \includegraphics[scale=.23]{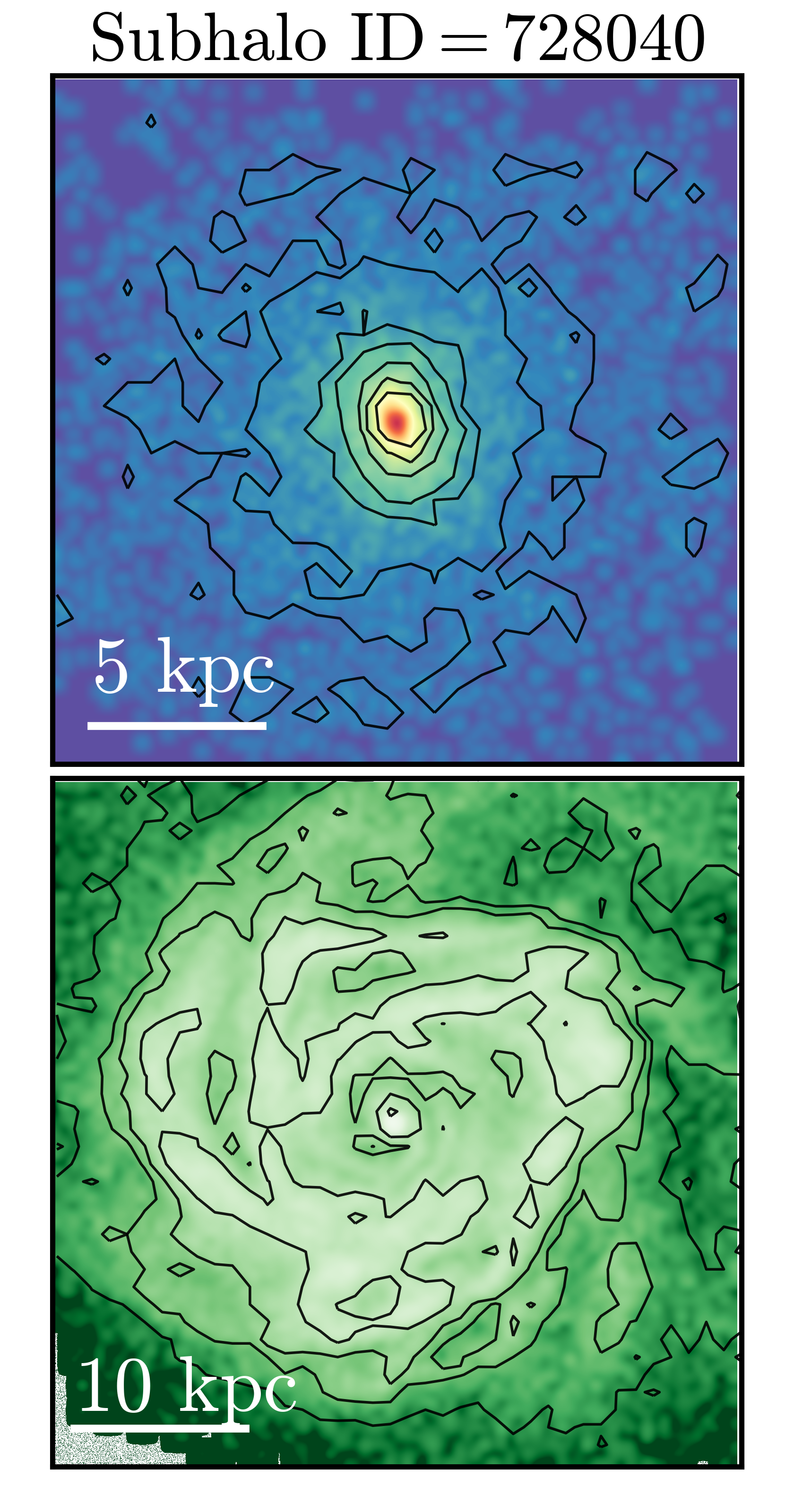}
\caption{UGC5288 analogue galaxies in TNG50 with their Subhalo IDs (698690, 712591, 721346, 728040) undergoing $\leq 1$ major merger all throughout their evolution. Among these galaxies, the first three galaxies (Subhalo ID- 698690, 721346 and 712591) have low halo spin similar to the halo spin of our predicted model with PISO halo profile and the last one (Subhalo ID-728040) have a higher halo spin. }
  \label{fig:analogues}
\end{figure}

We investigate halo spin distribution for barred and unbarred galaxies in the TNG50 \citep{Nelson.et.al.2019a,Pillepich.et.al.2019,Nelson.et.al.2019b} cosmological magneto hydrodynamical simulations at z=0. We find 24 similar stellar mass galaxy analogues of UGC5288 in the TNG50 simulations having similar flat rotation velocities, a small stellar disk with a weak bar ($A_{2}/A_{0}> 0.2$) and an extended gas disk. 75\% of the UGC5288 analogue sample ($\sim 18$ galaxies) have low halo spin similar to the halo spin prediction from our model and 4 out of the 24 galaxies undergo $\leq 1$ major merger event throughout their evolution (see Figure \ref{fig:analogues}). Hence a significant number of dwarf galaxies that undergo major mergers can still have low halo spin. Multiple merger properties, for example, satellite mass, number of pericenter passages, the inclination of orbit, and velocity of the incoming satellite will play important roles in determining the final halo spin of a galaxy at $z=0$. Currently, we have not considered the effect of mergers in our analysis.

\section{Discussion and conclusion}
We discuss the salient features of a new hybrid method to estimate the halo spin profile of an observed galaxy UGC5288 by forward modelling of the galaxy properties in N-body simulation models having a stellar disk, gas disk and a dark matter halo. We first need to constrain the DM halo profile from the HI rotation curve and next to match the bar properties of the disk in the models. The presence of a bar helps in breaking the degeneracy between different models with similar DM halo profiles. Additionally, the scatter in the values of halo spin is lower for barred galaxies than the unbarred galaxies. In the process of comparing our model predictions with similar stellar mass barred galaxies in TNG50, we find that it is difficult to find well-resolved dwarf galaxies having similar stellar mass and gas mass in the data set. For a more accurate comparison with the simulated UGC5288 models, we need high-resolution dwarf galaxy data set from cosmological simulations. To validate our method we plan to apply the method to the virtual galaxy data from cosmological simulations. In future work, we will improve the methodology by including star formation and the effect of mergers that are not included in the current modelling. 

\section*{Acknowledgement}
SA thanks the IIA HPC facility \textsf{NOVA} where the simulations are run. We thank Denis Yurin and Volker Springel for providing the codes \textsf{GalIC} and \textsf{GADGET2} publicly. We thank the IllustrisTNG team for making the data publicly available. MD acknowledges the support of the Science and Engineering Research Board (SERB) MATRICS grant MTR/2020/000266 for this research.

\end{document}